\documentclass[titlepage,oneside,fleqn,12pt]{article}

\pdfoutput=1

\usepackage[tiny,rm]{titlesec}
\newpagestyle{trbstyle}{
	\sethead{\small Roncoli, Bekiaris-Liberis, Papageorgiou}{}{\thepage}
}
\titleformat{\section}{\bfseries}{}{0pt}{\uppercase}
\titlespacing*{\section}{0pt}{12pt}{*0}
\titleformat{\subsection}{\bfseries}{}{0pt}{}
\titlespacing*{\subsection}{0pt}{12pt}{*0}
\titleformat{\subsubsection}{\itshape}{}{0pt}{}
\titlespacing*{\subsubsection}{0pt}{12pt}{*0}

\usepackage{enumitem}
\setlist[1]{labelindent=0.5in,leftmargin=*}
\setlist[2]{labelindent=0in,leftmargin=*}

\usepackage{ccaption}
\usepackage{amsmath}
\makeatletter
\renewcommand{\fnum@figure}{\textbf{FIGURE~\thefigure} }
\renewcommand{\fnum@table}{\textbf{TABLE~\thetable} }
\makeatother
\captiontitlefont{\bfseries \boldmath}
\captiondelim{\;}
\usepackage[sort,numbers]{natbib}
	\newcommand{\trbcite}[1]{\citeauthor{#1} ({\it \citenum{#1}})}
	\newcommand{\trbnum}[1]{{\it \citenum{#1}}}
\setcitestyle{round}

\makeatletter
	\renewcommand\@biblabel[1]{#1.\hspace*{15pt}}
\makeatother

\usepackage{mathptmx}

\usepackage[T1]{fontenc}
\usepackage{textcomp}

\usepackage{graphicx}
\usepackage[hidelinks]{hyperref}
\usepackage{color}
\usepackage{amsfonts}
\usepackage{multirow}

\begin{document}

\thispagestyle{empty}

\begin{titlepage}
\begin{flushleft}

{\bfseries HIGHWAY TRAFFIC STATE ESTIMATION USING SPEED MEASUREMENTS: CASE STUDIES ON NGSIM DATA AND HIGHWAY A20 IN THE NETHERLANDS}\\[1.5cm]

\textbf{Claudio Roncoli, Corresponding Author} \\
Dynamic Systems and Simulation Laboratory \\
Technical University of Crete \\
Chania, 73100, Greece \\
Tel: +30 28210 37289; Fax: +30 28210 37584; Email: \href{mailto:croncoli@dssl.tuc.gr}{\textcolor{blue}{\underline{\smash{croncoli@dssl.tuc.gr}}}}\\[0.5cm]

\textbf{Nikolaos Bekiaris-Liberis} \\
Dynamic Systems and Simulation Laboratory \\
Technical University of Crete \\
Chania, 73100, Greece \\
Email: \href{mailto:nikos.bekiaris@gmail.com}{\textcolor{blue}{\underline{\smash{nikos.bekiaris@gmail.com}}}}\\[0.5cm]

\textbf{Markos Papageorgiou} \\
Dynamic Systems and Simulation Laboratory \\
Technical University of Crete \\
Chania, 73100, Greece \\
Email: \href{mailto:markos@dssl.tuc.gr}{\textcolor{blue}{\underline{\smash{markos@dssl.tuc.gr}}}}\\[1cm]




\vspace{2.5cm}

\today
\end{flushleft}
\end{titlepage}

\newpage

\setcounter{page}{2}

\section{Abstract}
This paper presents two case studies where a macroscopic model-based approach for traffic state estimation, which we have recently developed, is employed and tested. The estimation methodology is developed for a ``mixed'' traffic scenario, where traffic is composed of both ordinary and connected vehicles. Only average speed measurements, which may be obtained from connected vehicles reports, and a minimum number (sufficient to guarantee observability) of spot sensor-based total flow measurements are utilised.
In the first case study, we use NGSIM microscopic data in order to test the capability of estimating the traffic state on the basis of aggregated information retrieved from moving vehicles and considering various penetration rates of connected vehicles. In the second case study, a longer highway stretch with internal congestion is utilised, in order to test the capability of the proposed estimation scheme to produce appropriate estimates for varying traffic conditions on long stretches. In both cases the performances are satisfactory, and the obtained results demonstrate the effectiveness of the methodology, both in qualitative and quantitative terms.
\\[2.5cm]
\textit{Keywords:} traffic state estimation, connected vehicles, Kalman filter, real-data testing
\newpage

\section{Introduction} \label{sec:intro}

Traffic state estimation is a task of crucial importance for the development and application of traffic management and control strategies. Essentially, it consists in estimating the traffic variables of a highway network at sufficient spatial resolution in real-time based on limited amount of information available from real-time traffic measurements. In conventional traffic, the necessary measurements are provided by spot sensors (based on a variety of possible technologies), which are placed at specific highway locations. In case the sensor density is sufficiently high (e.g., every 500 m), the collected measurements are usually sufficient for traffic surveillance and control; otherwise, appropriate estimation schemes need to be employed in order to produce traffic state estimates at the required space resolution; see, for instance, (\trbnum{Munoz2003}, \trbnum{Wang2005}, \trbnum{Mihaylova2007}, \trbnum{Morbidi2014}) among many other works addressing the problem of highway traffic estimation employing conventional detector data.

In the last two decades there has been a significant and increasing interdisciplinary effort by the automotive industry as well as by numerous research institutions around the world to plan, develop, test and start deploying a variety of Vehicle Automation and Communication Systems (VACS) that are expected to revolutionise the features and capabilities of individual vehicles within the next decades. A wide description of VACS, focusing particularly on their impact on highway traffic efficiency, may be found in (\trbnum{Diakaki2015}).
To achieve related traffic flow efficiency improvements on highways, it is of paramount importance to develop novel methodologies for modelling, estimation and control of traffic in presence of VACS. Several papers are providing interesting results related to traffic control in presence of VACS, employing either microscopic or macroscopic approaches; see, for example (\trbnum{Varaiya1993}, \trbnum{Rao1994a}, \trbnum{Rajamani2001}, \trbnum{Bose2003a}, \trbnum{Bose2003b}, \trbnum{vanArem2006}, \trbnum{Kesting2008}, \trbnum{Shladover2012}, \trbnum{Ge2014}, \trbnum{Wang2014}, \trbnum{Roncoli2015c}, \trbnum{Roncoli2014d}).

Some VACS are characterised by different types of communication devices, providing the possibility of V2V (vehicle-to vehicle) or V2I (vehicle-to-infrastructure) communication; specifically, ``connected'' vehicles can report their position, speed, and other relevant information, i.e., they can act as mobile sensors; existing examples include mobile phones and network--connected GPS (Global Positioning System) devices. From a traffic estimation viewpoint, this may allow for a sensible improvement of the achievable estimation accuracy, as well as significant reduction of the necessary number of spot sensors; the latter would lead to a sensible reduction of the purchase, installation, and maintenance cost of traffic monitoring.

The exploitation of information retrieved from connected vehicles for traffic estimation purposes is considered in numerous research works. \trbcite{Work2008} designed an ensemble Kalman filter for a Godunov-discretised Lighthill-Whitham-Richards (LWR) model, estimating the velocity field on highways using data obtained from GPS devices; a real-case test is defined based on data retrieved from the Mobile Century experiment (\trbnum{Herrera2010b}). The work by \trbcite{Treiber2011} is based on data fusion techniques, utilising information retrieved both from fixed and mobile sensors; two case studies are reported, based on data from the highway A9, near Munich, Germany, and from highway M42, near Birmingham, UK. \trbcite{Herrera2010a} designed a methodology to incorporate mobile probe measurements into a modified LWR model, exploiting the so-called Newtonian relaxation; the efficiency and performance of the estimation algorithm were tested on highway I-880 in California within the frame of the Mobile Century experiment (\trbnum{Herrera2010b}). Yuan et al. (\trbnum{Yuan2012}) formulated a traffic model in a Lagrangian coordinate system (where state variables move with the traffic stream), including both Eulerian and Lagrangian measurements; an extended Kalman filter is then used to obtain appropriate traffic estimates; and the proposed method is tested on data retrieved from highway M42, near Birmingham, UK. \trbcite{Piccoli2015} addressed the impact of both sampling frequency and penetration rate on mobile sensing of highway traffic, employing a second-order phase transition model (PTM) and proposing several data fusion schemes to incorporate vehicle trajectory data into the PTM; extensive tests using NGSIM data (\trbnum{Ngsim}) are performed. \trbcite{Seo2015a} proposed a novel probe vehicle-based estimation method for obtaining volume-related variables by assuming that a probe vehicle can measure the spacing to its leading one; remarkably, they do not assume the knowledge of any fundamental diagram and perform a real experiment employing prototypes of probe vehicles. This method is extended and improved by \trbcite{Seo2015b} taking into consideration the conservation law. \trbcite{Bekiaris2015} proposed a macroscopic model-based approach for the estimation of the total density and total flow; i.e., the density/flow of all vehicles (connected and ordinary), for the case of ``mixed'' traffic utilising speed measurements that may be reported by connected vehicles, and a small number of spot flow measurements; this approach removes the requirement of (empirical, hence uncertain) traffic speed modelling, such as the fundamental diagram. The performance of the developed estimation schemes was validated through simulation using the well-known second-order traffic flow model METANET (\trbnum{Papageorgiou1990}) as ground truth.

In this paper, we present two case studies using real data retrieved from highways in order to test and evaluate the behaviour of the recently developed estimation scheme (\trbnum{Bekiaris2015}).
Specifically, we address the problem of estimating the (total) density and flow of vehicles in highway segments of arbitrary length in presence of connected vehicles.
The two case-studies are built using data stemming from different real experiments. The first case, based on the NGSIM microscopic data, aims at verifying the capability to estimate the traffic state at high space-resolution (i.e., considering segments in the order of 50\,m in length), using aggregate information retrieved from moving vehicles, that is, the speeds of connected vehicles and the total flow measured at the entrance and exit of the highway stretch; the proposed estimation scheme is also tested for different penetration rates of connected vehicles. In the second case, a longer highway stretch and a longer time horizon are considered, using information coming from fixed detectors, namely speed measurements at every segment (emulating connected vehicles reports) and a small amount of flow measurements. This permits to verify the capability of detecting the onset of a congestion created within the network, which is one of the most crucial and challenging aspects in case this approach is exploited in conjunction with a traffic control strategy. In both cases, the estimation results are analysed qualitatively as well as quantitatively, employing an appropriate metric to compare the performance of the proposed methodology under different scenarios.

In the following section, we present a linear time-varying model of the density dynamics and a Kalman filter for density estimation; then two case studies to evaluate the performance of the estimation scheme are described; finally, the paper is concluded highlighting the main findings and introducing further challenges for future research.

\section{Traffic Estimation Using Average Speed Measurements} \label{sec:filter}

\subsection{Traffic Density Dynamics} \label{sec:model}
We consider the following discrete-time equations that describe the dynamics of the total densities $\rho_i$, namely the number of vehicles within segment $i$ divided by the the length of the segment $\Delta_i$, on subsequent highway segments (see, e.g., (\trbnum{Papageorgiou1990a}))
\begin{equation}\label{eqrho}
\rho_i(k+1) = \rho_i(k)+\frac{T}{\Delta_i}\big(q_{i-1}(k)-q_i(k)+r_i(k)-s_i(k)\big),
\end{equation}
where $i=1,\ldots,N$ is the index of the specific highway segment, $N$ being the number of segments on the highway. For all traffic variables, we denote by index sub-$i$ its value at the segment $i$ of the highway; $q_i$ is the total flow at the exit of segment $i$; $T$ is the time-discretisation step; and $k=0,1,\ldots$ is the discrete time index. The variables $r_i$ and $s_i$ denote the inflow and outflow of vehicles at on-ramps and off-ramps, respectively, at segment $i$. Using the known relation
\begin{equation}\label{flowtotal}
q_i(k) = \rho_i(k) v_i(k),
\end{equation}
where $v_i$ is the average speed in segment $i$, we write Equation \ref{eqrho} as
\begin{equation}\label{eqrh1}
\rho_i(k+1) = \frac{T}{\Delta_i}v_{i-1}(k)\rho_{i-1}(k)+\left(1-\frac{T}{\Delta_i}v_i(k)\right)\rho_i(k)+\frac{T}{\Delta_i}\left(r_i(k)-s_i(k)\right).
\end{equation}
Note that, in order to guarantee that the discrete-time relation described by Equation \ref{flowtotal} is sufficiently accurate, the following inequality must be satisfied
\begin{equation}\label{eqStab}
\max_{i,k} \frac{\Delta_i}{T} v_{i}(k) < 1.
\end{equation}

Assuming that the unmeasured on-ramp and off-ramp flows are constant (or, effectively, slowly varying), the unmeasured ramp flow dynamics may be reflected by a random walk, i.e., 
\begin{eqnarray}
\theta_i(k+1)=\theta_i(k)+\xi_i^{\theta}(k),\label{thetai}
\end{eqnarray}
 where $\xi_i^{\theta}$ is zero-mean white Gaussian noise and 
 \begin{eqnarray}
 \theta_i=\left\{\begin{array}{ll}\frac{T}{\Delta_i}r_{n_i},&\mbox{if $n_i\in L_r$}\\\frac{T}{\Delta_i}s_{n_i},&\mbox{if $n_i\in L_s$}\end{array}\right\},
\end{eqnarray} 
for all $i=1,\ldots,l_r+l_s$, with $L_r=\left\{n_1,\ldots,n_{l_r}\right\}$ and $L_s=\left\{n_{l_r+1},\ldots,n_{l_r+l_s}\right\}$ being the sets of segments, denoted by $n_i$, which have an on-ramp or an off-ramp, respectively, whose flows are not directly measured; and $l_r$ and $l_s$ are the number of unmeasured flows at on-ramps and off-ramps, respectively. We assume that at a segment $i$ there can be either only one on-ramp or only one off-ramp, which is typically the case on a highway, and hence, $L_r\cap L_s=\emptyset$.

Assuming that the average speed of vehicles in a segment, namely $v_i$, is measured (e.g. from connected vehicle reports) and defining the state
\begin{eqnarray}
{x}=\left(\rho_1,\ldots,\rho_N,\theta_1,\ldots,\theta_{l_r+l_s}\right)^T, \label{stnew}
\end{eqnarray}
the deterministic part of the dynamics of the total density given in Equation \ref{eqrh1} and of $\theta_i$ given in Equation \ref{thetai} can be written in the form of a linear time-varying system as
\begin{eqnarray}
{x}(k+1)={A}(k){x}(k)+{B}{u}(k),\label{barx}
\end{eqnarray}
where
\begin{eqnarray}
{A}(k)&=&\left\{\begin{array}{lll}{a}_{ij}=\frac{T}{\Delta_i}v_{i-1}(k),&\mbox{if $i-j=1$ and $i\geq2$}\\{a}_{ij}=1-\frac{T}{\Delta_i}v_i(k),&\mbox{if $i=j$}\\{a}_{n_ij}=1,&\mbox{if $n_i\in L_r$ and $j=N+i$}\\{a}_{n_ij}=-1,&\mbox{if $n_i\in L_s$ and $j=N+i$}\\{a}_{ij}=1,&\mbox{if $N< i\leq N_1$ and $j=i$}\\{a}_{ij}=0,&\mbox{otherwise}\end{array}\right\}\label{adef1}\\
{B}\!&=&\!\left[\begin{array}{ll}{b}_{ij}=\frac{T}{\Delta_i},&\mbox{if $i=1$ and $j=1$}\\{b}_{m_ij}=\frac{T}{\Delta_{m_i}},&\mbox{if $m_i\notin \bar{L}$, $1\leq m_i\leq N$, $1\leq i\leq N_2$, and $j=i+1$}\\{b}_{ij}=0,&\mbox{otherwise}\end{array}\!\!\right]\label{defbu}\\
{u}(k)&=&\left[\begin{array}{ll}{u}_i=q_0(k),&\mbox{if $i=1$}\\{u}_{i+1}=r_{m_i}-s_{m_{i}},&\mbox{if $m_i\notin \bar{L}$}\end{array}\right],\label{newu}
\end{eqnarray}
with $\bar{L}=L_r\cup L_s$, $N_1=N+l_r+l_s$, $N_2=N-l_r-l_s$, ${A}\in\mathbb{R}^{N_1\times N_1}$, ${B}\in\mathbb{R}^{N_1\times (N_2+1)}$. Note that $q_0$, which is assumed to be measured (for example, via fixed flow detector), denotes the total flow of vehicles at the entry of the considered highway stretch and acts as an input to Equation \ref{barx}; along with any directly measured on-ramp and off-ramp flows, $r_i$ and $s_i$, $i \notin L_r$, $i \notin L_s$ respectively; while $v_i$, $i=1,\ldots,N$, are viewed as time-varying parameters of Equation \ref{barx}.

We now turn our attention to the measured outputs. We assume availability of the density (or, equivalently, of the flow) at the mainstream exit of the highway. If there is exactly one unmeasured ramp within the considered highway stretch, then no additional measurements are necessary for flow observability. On the other hand, if there are more than one unmeasured ramps within the stretch, we need one mainstream flow measurement at any highway segment between every two consecutive unmeasured ramps.  

In summary, the measured outputs associated with Equations \ref{barx}--\ref{newu} are the density (or, equivalently, the flow) at the exit of the considered highway stretch and at a highway segment between every two consecutive ramps whose flows are not measured. Therefore,
\begin{eqnarray}
{y}(k)={C}{x}(k),\label{newy}
\end{eqnarray} 
where ${C}\in\mathbb{R}^{(l_r+l_s)\times (N+l_r+l_s)}$ is defined as
\begin{eqnarray}
{C}&=&{\left[\begin{array}{ll}{c}_{ij}=1,&\mbox{for all $i=1,\ldots,l_r+l_s-1$ and some $n_i^*\leq j\leq n_{i+1}^*-1$}\\{c}_{ij}=1,&\mbox{if $i=l_r+l_s$ and $j=N$}\\{c}_{ij}=0,&\mbox{otherwise}\end{array}\right]},\label{16rhonew}
\end{eqnarray}
where $\bar{L}^*=\left\{n_1^*,n_2^*,\ldots,n_{l_r+l_s}^*\right\}$ is the set $\bar{L}$ ordered by $<$.

Although it is physically intuitive that the system described in Equations \ref{barx}--\ref{16rhonew} is observable, when the additional mainstream fixed flow sensors are placed at some segment between every two consecutive unmeasured ramps, we rigorously proved in (\trbnum{Bekiaris2015}) that the system is indeed observable for certain cases, such as, for example, when a fixed sensor is placed on the mainstream at every segment immediately before an unmeasured ramp.

We summarise below the measurement requirements for the proposed estimation algorithm.
\begin{itemize}
\item The average speed of connected vehicles at a segment of the highway, $v_i$, $i=1,\ldots,N$, are measured; the assumption being that there is no systematic  difference between the average speed of connected vehicles and the average speed of all vehicles in a segment.
\item The total flow of vehicles at the entry and exit of the considered highway stretch, $q_0$ and $q_N$, respectively.
\item Either the total flow of vehicles at a ramp, say, $r_i$ or $s_i$, is measured or an additional mainstream flow measurement, say $q_j$, at any highway segment between two consecutive unmeasured ramps, is available.
\end{itemize}

\subsection{Kalman Filter} \label{sec:kal2}
\begin{figure}
\begin{center}
	\includegraphics[width=0.7\textwidth]{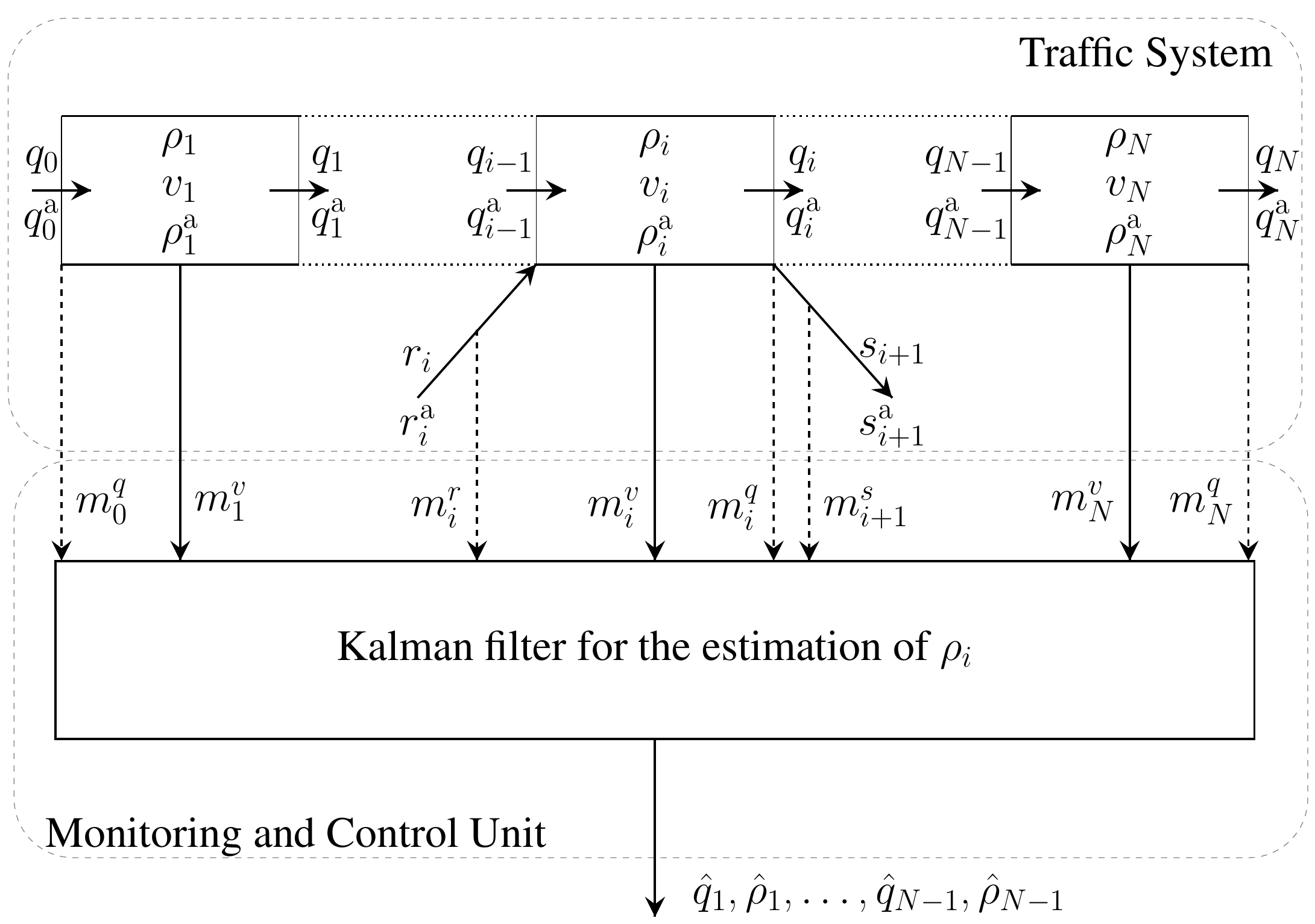}
	\caption{The traffic system under consideration and the proposed Kalman filter implementation. The data used to operate the Kalman filter are speed measurements coming from connected vehicles (solid lines) and flow measurements coming from fixed sensors (dashed lines). The variable $m_i^w$ denotes the measurement of quantity $w$ at segment $i$, which might be different than the actual quantity $w$, due to, for example, the presence of measurement noise. A variable $w_i^{\rm a}$ represents the value of quantity $w$ of connected vehicles, which may be present, at segment $i$.}
	\label{fig1}
\end{center}
\end{figure}

We employ a Kalman filter for the estimation of the total density of vehicles in each segment of a highway (Figure \ref{fig1}). Defining the estimated state vector
\begin{eqnarray}
\hat{{x}}=\left(\hat{\rho}_1,\ldots,\hat{\rho}_N,\hat{\theta}_1,\ldots,\hat{\theta}_{l_r+l_s}\right)^T,
\end{eqnarray}
the Kalman filter equations are (see e.g., (\trbnum{Anderson1979}))

\begin{eqnarray}
\hat{{x}}(k+1)&=&{A}(k)\hat{{x}}(k)+{B}{u}(k)+{{A}}(k){K}(k)\left({{z}}(k)-{C}\hat{{x}}(k)\right)\label{123}\\
{K}(k)&=&{P}(k){C}^T\left({C}{P}(k){C}^T+{R}\right)^{-1}\\
{P}(k+1)&=&{A}(k)\left(I-{K}(k){C}\right){P}(k){A}(k)^T+{Q},\label{kal1}
\end{eqnarray}
where ${z}$ is a noisy version of the measurement ${y}$, ${R}=R^T>0$ and ${Q}={Q}^T>0$ are tuning parameters. Note that, in the ideal case in which there is additive, zero-mean Gaussian white noise in Equations \ref{barx} and \ref{newy}, respectively, $R$ and $Q$ represent the (ideally known) covariance matrices of the measurement and process noises, respectively (\textcolor{black}{in particular, the ${q}_{N+iN+i}$ elements of ${Q}$ represent the filter's anticipation for the covariance of $\xi_i^{\theta}$}). Since the system equations here are relatively complex, some tuning of $R$, $Q$ may be needed for best estimation results.
The initial conditions of the filter described by Equations \ref{123}--\ref{kal1} are chosen as
\begin{eqnarray}
\hat{{x}}(k_0)&=&\mu\\
{P}(k_0)&=&H,\label{1234}
\end{eqnarray}
where $\mu$ and $H=H^T>0$, which, in the ideal case in which ${x}(k_0)$ is a Gaussian random variable, represent the mean and auto covariance matrix of ${x}(k_0)$, respectively. The Kalman filter delivers estimates of the segment total densities $\hat{\rho}_i$, as indicated at the output of the Kalman filter in Figure~\ref{fig1}.

\section{Experimental testing of the proposed estimation methodology} \label{sec:experiment}

\subsection{Case Study 1: NGSIM I-80 Data}

\subsubsection{Network and Data Description}

\begin{figure}
\begin{center}
	\includegraphics[width=0.8\textwidth]{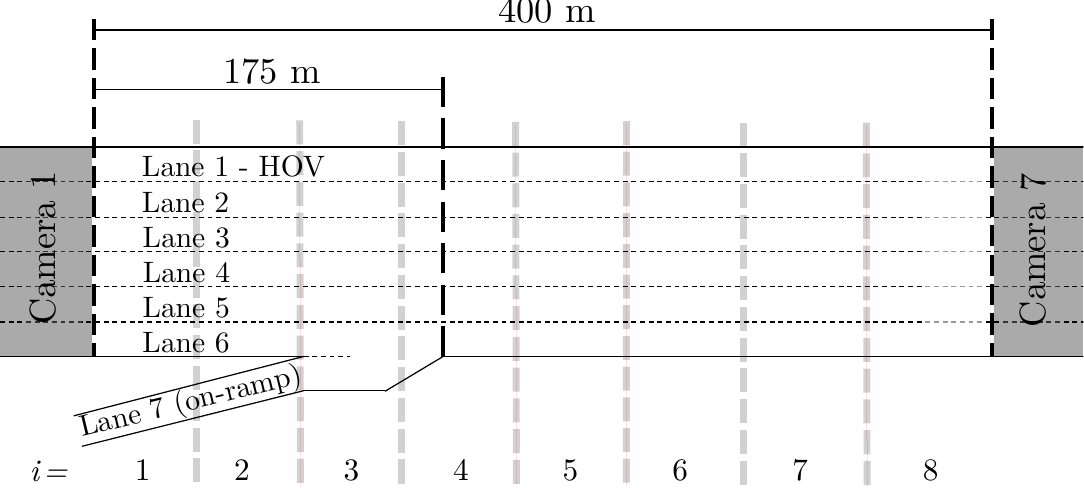}
	\caption{A graphical representation of the stretch of the highway I-80 in Emeryville, California, used in Case Study 1.} 
	\label{fig:ngsimStretch}
\end{center}
\end{figure}

In order to evaluate and illustrate the effectiveness of the proposed estimation scheme, microscopic traffic data collected within the Next Generation SIMulation program (\trbnum{Ngsim}) are utilised. However, since these data incorporate non-negligible errors in the position of individual vehicles (\trbnum{Punzo2011}), some correction methodologies were proposed to improve their reliability (\trbnum{Montanino2013}, \trbnum{Piccoli2015}). This work utilises the data processed by Montanino and Punzo (\trbnum{Montanino2013}, \trbnum{NgsimPunzo}), which include the trajectories of all vehicles travelling along a stretch in the northbound direction of I-80 in Emeryville, California, recorded from 4:00 PM. to 4:15 PM on April 13, 2005. The highway is composed of 6 lanes; however, lane 1 is a so-called HOV (high-occupancy vehicle) lane, characterised by access restricted to a limited set of vehicles (only vehicles with a driver and one or more passengers are allowed), causing its traffic characteristics to be structurally different compared with the other lanes; therefore this lane is excluded from our estimation.
The considered stretch (sketched in Figure \ref{fig:ngsimStretch}) is 400\;m long and an on-ramp is entering the mainstream, where the merge nose is located 175\,m after the network origin.

\begin{figure}
\begin{center}
	\includegraphics[width=0.7\textwidth]{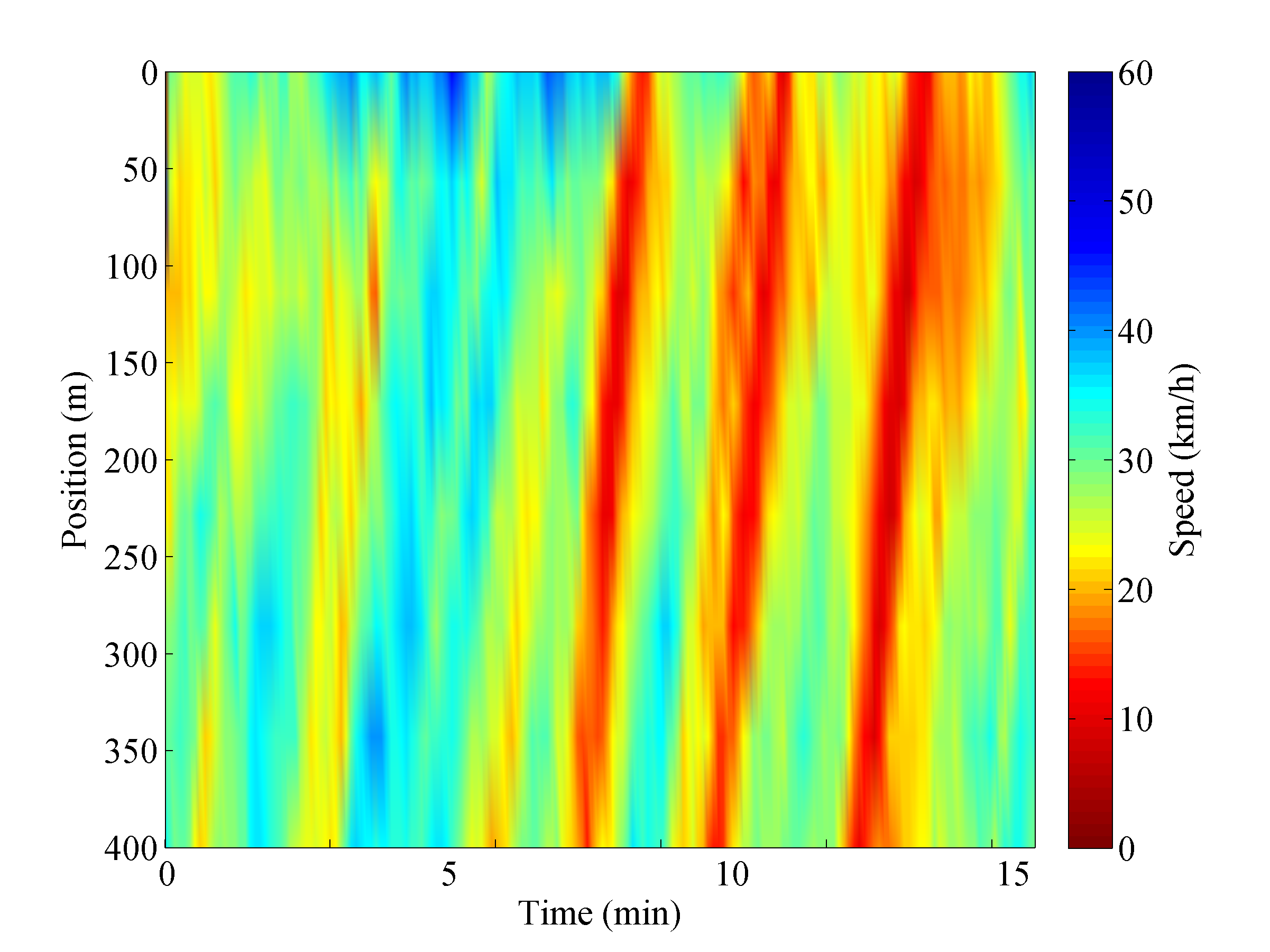}
	\caption{The aggregated vehicle speeds extracted from the NGSIM data. Congestion shock-waves are spilling back from downstream and cross the entire highway stretch.} 
	\label{fig:ngsim3d}
\end{center}
\end{figure}

As it can be seen from Figure \ref{fig:ngsim3d}, a massive congestion is present within the stretch, where congestion waves are coming from downstream and crossing the entire stretch.
In order to perform macroscopic evaluations, the stretch is divided into $N=8$ homogeneous segments of 50\,m in length (see Figure \ref{fig:ngsimStretch}); the on-ramp is placed within segment $4$; while a discrete time step $T=5\,$s is used. Even though this configuration slightly violates the condition of Equation \ref{eqStab} for some time intervals and some segments, the obtained results were found to be very similar to those obtained when using lower values for the time step $T$, which satisfy fully the condition of Equation \ref{eqStab}.

All necessary information to perform estimation is extracted from the available trajectory data. In particular, the segment speed $v_i(k)$ is computed as the arithmetic average of the speeds of connected vehicles present within the segment at the measurement time instant $kT$; whereas total flow measurements are computed via two virtual spot detectors; one placed at the network entrance (beginning of segment $i=1$), providing $q_0$, and another placed at the network exit (end of segment $i=8$), providing $q_N$; the corresponding flow values are computed by counting the number of vehicles (irrespectively of being connected or not) that cross the virtual detector locations during the corresponding period.

The estimation scheme is configured in order to estimate the mainstream densities and the total on-ramp flow. In the first scenario, we suppose that all vehicles are connected, thus the segment speed is computed using information from all vehicles present; whereas in the second scenario, variable percentages of connected vehicles are assumed to be present in the highway, thus the estimation algorithm uses a less accurate measurement of the segment speeds. In the latter case, vehicles entering the network are randomly marked as connected, according to the assumed penetration rate and based on a uniform distribution.

The ground truth, used for evaluating the estimation results, is represented by the total density in each segment and the on-ramp flow. The segment densities $\rho_i(k)$ are computed by counting the number of vehicles that are present within segment $i$ at time instant $kT$ divided by the segment length (0.05\,km); whereas, the on-ramp flow $r_4(k)$, is computed similarly to the computation of flow by a virtual detector, i.e. by counting the number of vehicles leaving the on-ramp (Lane 7 in Figure \ref{fig:ngsim3d}) and entering the mainstream during time interval $(k,k+1]$.

\subsubsection{Performance Evaluation Under 100\% Penetration Rate}

In the first scenario, it is assumed that all vehicles are connected, thus capable of communicating their speed and position to the central authority; consequently, the filter is fed with accurate measurements of the average segment speeds. The parameters used for the Kalman filter are the following: $ Q_{i,i}=1, \; 1 \leq i \leq N; Q_{i,i}=0.01, \; N < i \leq N+l_r+l_s; Q_{i,j}=0, \; \forall \; i \neq j$;  $R=10 \; I_{ \left( l_r+l_s \right) \times \left( l_r+l_s \right) }$; $\mu = \left(40,\ldots,40\right)$; $H=I_{ \left(N+l_r+l_s\right) \times \left(N+l_r+l_s\right) }$.

It should be noted that under 100\% penetration rate, the filter would not be really needed, because the total densities and flows could be directly calculated based on vehicle position report. Nevertheless, we apply the estimation scheme also for this case to enable comparison with lower penetration-rate results.

The following performance index, formulated as the Coefficient of Variation (CV) of the root mean square error of the estimated density $\hat{\rho}$ with respect to the ground truth density $\rho$, is proposed for numerical evaluation of the estimation methodology:
\begin{equation} \label{eq:perfIndexDens}
CV_\rho = \frac{\sqrt{ \frac{1}{KN} \sum_{k=1}^{K} \sum_{i=1}^{N} \left[ \hat{\rho}_i(k) - \rho_i(k) \right]^2}}{\frac{1}{KN} \sum_{k=1}^{K} \sum_{i=1}^{N} \rho_i(k)}.
\end{equation}

\begin{figure}
\begin{center}
	\includegraphics[width=\textwidth]{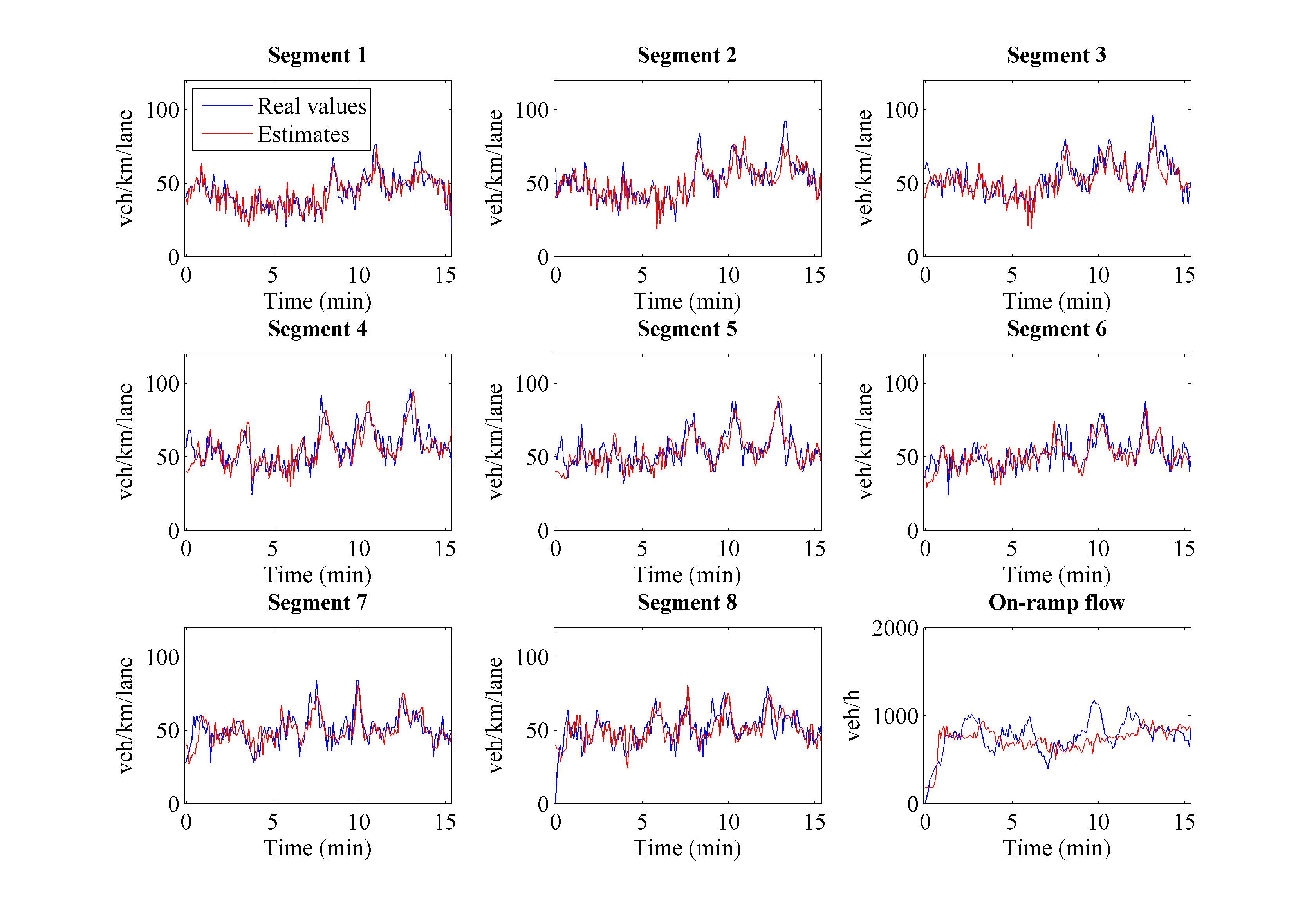}
	\caption{The trajectories of real and estimated densities for the NGSIM case study, assuming that all vehicles are connected (perfect knowledge of segment speeds).} 
	\label{fig:traj100}
\end{center}
\end{figure}

The obtained results are characterised by $CV_\rho=14.9\%$, which imply a very accurate estimation; this is also visible in Figure \ref{fig:traj100}, where it is clear that the produced estimates capture with high accuracy the dynamic variations of traffic states.
The estimation of the ramp flow suffers from some delay that is caused by the distance of the downstream detector from the on-ramp location; in fact, the dynamics of flow entering from the on-ramp can be captured only after the on-ramp vehicles reach the measurement location (i.e., at the end of the stretch). As a matter of fact, these discrepancies may be substantially reduced by placing an additional flow detector immediately downstream of the merging segment.

\subsubsection{Performance Evaluation Under Various Penetration Rates}

Within corresponding experiments we consider mixed traffic (of connected and ordinary vehicles) with different penetration rates of connected vehicles. Essentially, this corresponds to evaluating the robustness of the approach to potentially less accurate measurements of the segment speeds. In this case, because of the limited number of vehicles transmitting information, two main issues may arise: first, the measurement retrieved from a limited number of vehicles may not be fully representative for the real average speed of a segment; and, second, there may be no connected vehicles within some segments when the measurements are retrieved. Potential reasons for the former are related to microscopic and behavioural phenomena; for example, consider the case that there is only one connected vehicle in a segment and its speed is retrieved after the driver is forced to strongly decelerate because of an aggressive lane-changing manoeuvre from some other vehicle, or the connected vehicle is driven by a particularly fast (or slow) driver. To compensate for potential large errors in the reported segment speeds in such situations, a moving average of the reported segment speeds is employed for feeding the filter, rather than using instantaneous segment speeds. In particular, the last three measurements are averaged (thus, considering a time window of 10 s), ignoring measurements at time steps when no speed is reported (because no connected vehicles are present).

\begin{figure}
\begin{center}
	\includegraphics[width=0.8\textwidth]{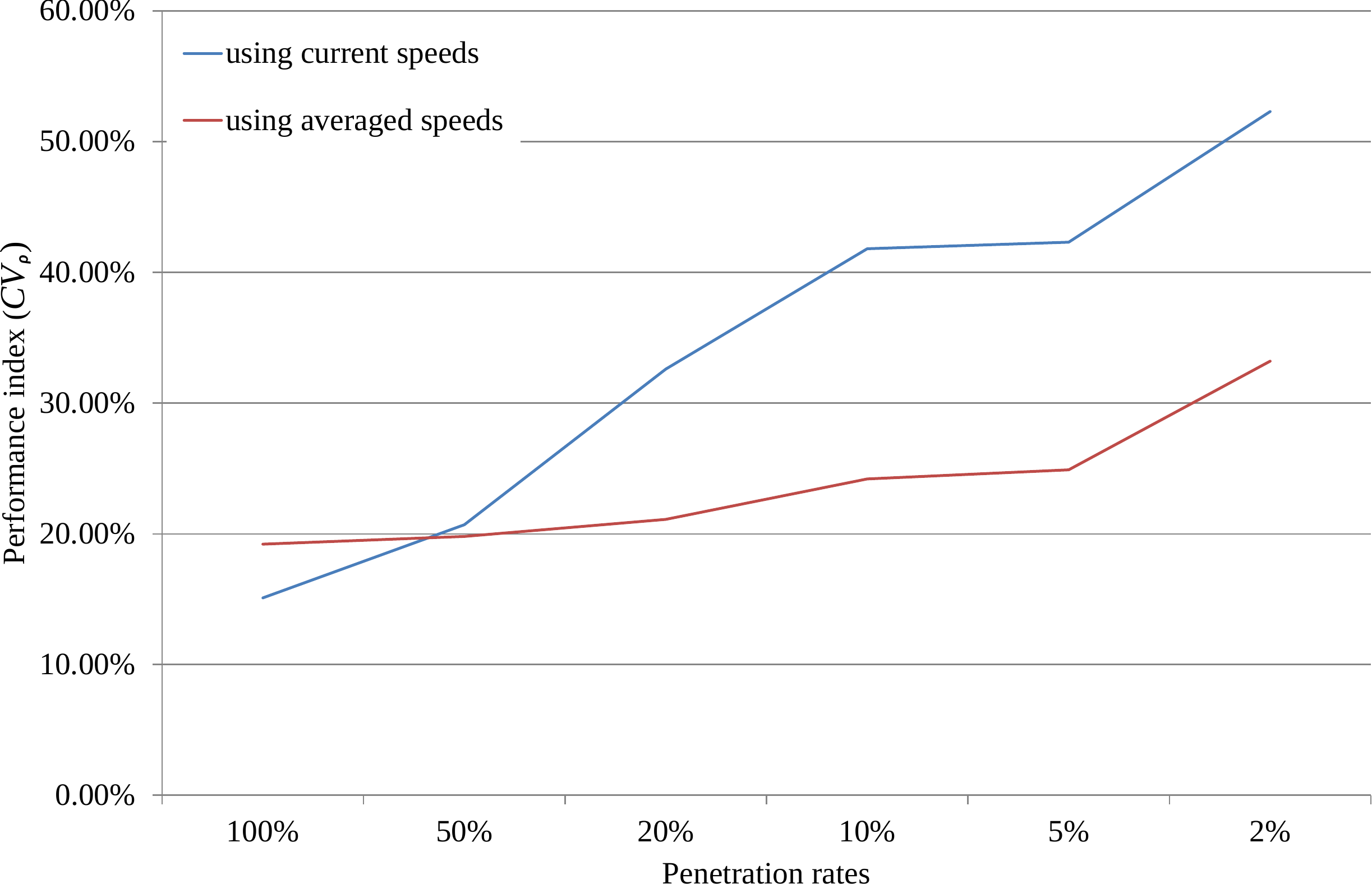}
	\caption{Comparison of performance indexes of the relative error for the NGSIM data, considering different penetration rates of connected vehicles, using the current speeds and a moving average of the reported speeds during the last 10\,s.} 
	\label{fig:penRates}
\end{center}
\end{figure}

\begin{figure}
\begin{center}
	\includegraphics[width=\textwidth]{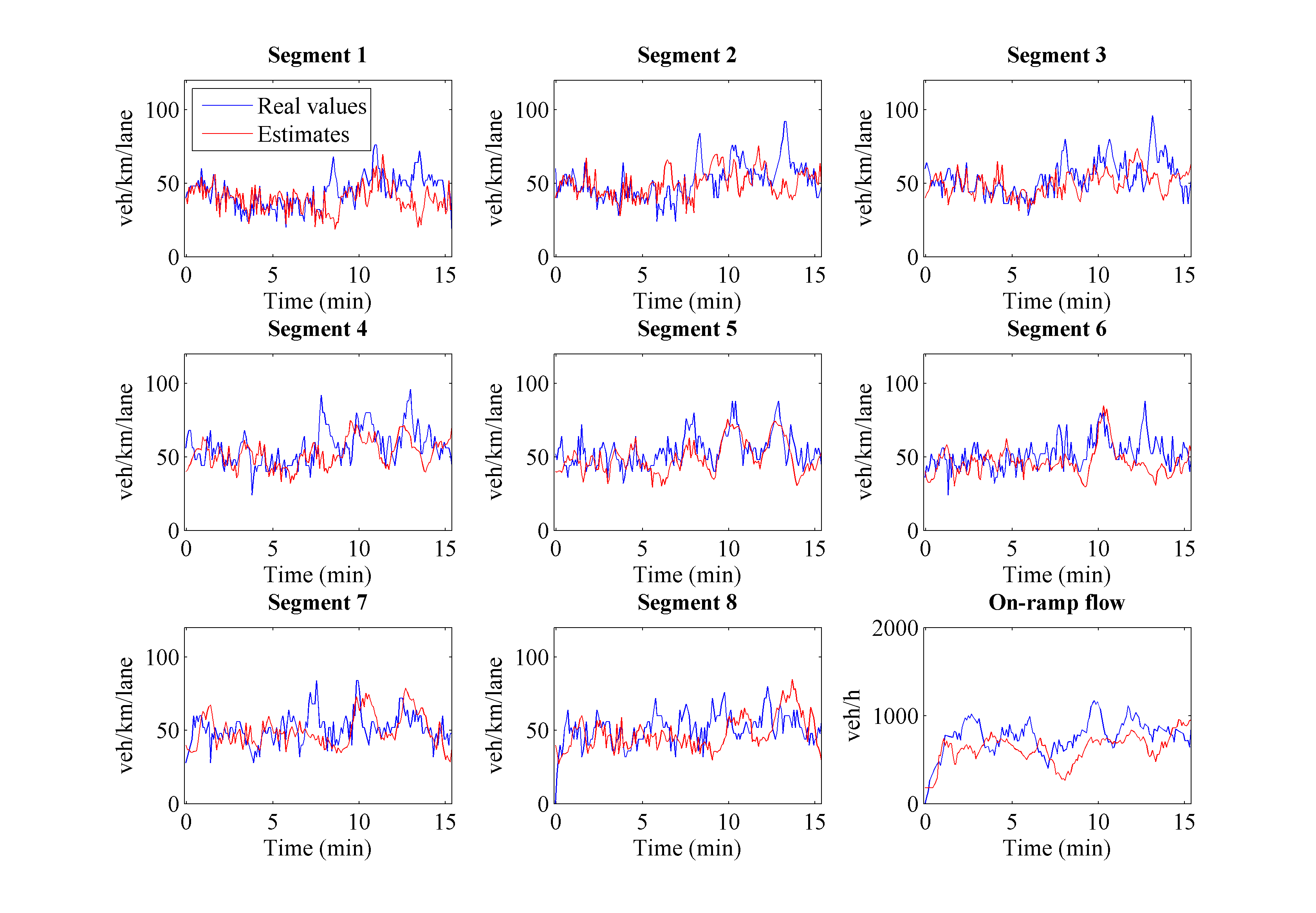}
	\caption{The trajectories of real and estimated densities for the NGSIM case study, assuming a 5\% penetration rate of connected vehicles; the filter is fed with a moving average of the reported speeds during the last 10\,s.} 
	\label{fig:traj5}
\end{center}
\end{figure}

The parameters used for the Kalman filter are kept the same as in the previous case.
From the performance index comparison displayed in Figure \ref{fig:penRates}, it appears that the use of a moving average for the reported segment speeds is indeed advantageous over the use of instantaneous speeds for low penetration rates and that the performance of the estimation algorithm slowly degrades with respect to a decrease of penetration rates, with the performance index $CV_\rho$ never reaching 35\% in all tested cases.
Thus, the estimation scheme is seen to produce good results even if the penetration rate of connected vehicles is very low (e.g. 2\%).
For the sake of brevity, only the results related to one of these tests, namely with a 5\% penetration rate of connected vehicles and when feeding the filter with the moving average of the reported speeds during the last 10\,s, is illustrated in Figure \ref{fig:traj5}; despite the obviously less accurate estimation, compared to Figure \ref{fig:traj100}, we can observe that the estimator is capable of following the real density pattern and the on-ramp flow dynamics (with some delay).

We performed also some tests in order to evaluate the magnitude of the error related to the incorrect measurement of average segment speeds for some cases where low penetration rate of connected vehicles is considered. In particular, we run 10 simulations considering different sets of vehicles being connected (for a fixed penetration rate), computing then the mean covariance $w$ of the related conservation equation error due to discrepancy between the segment speed computed using information from connected vehicles $\hat{v}_i(k)$ and the real average segment speed $\bar{v}_i(k)$, according to
\begin{equation} \label{covar}
w = \frac{1}{KN} \sum_{i=1}^N \sum_{k=1}^K \frac{T^2}{\Delta_i^2} \; E \left[ \rho_i^2(k) \left( \hat{v}_i(k) - \bar{v}_i(k) \right)^2 \right] .
\end{equation}
The resulting value for a penetration rate of connected vehicles equal to 20\% is $w=8.1\,\mbox{veh}^2/\mbox{km}^2$, while for a penetration rate of 5\% the resulting mean covariance is $w=19.2\,\mbox{veh}^2/\mbox{km}^2$. These values suggest that better performance may be achieved by utilising higher values of $Q$ for lower penetration rates. Nevertheless, experimental tests not reported here) indicate that, for low penetration rates, the performance of the proposed estimation scheme does not vary significantly for different values of $Q$.

Unfortunately, NGSIM data were collected for a short highway stretch within a peak period characterised by congested traffic conditions only; therefore the capability of the proposed estimation scheme in capturing the flow breakdown occurrence cannot be tested with these data. In order to extend the evaluation range of the proposed estimation scheme, another case study is performed and it is presented in the next section.

\subsection{Case Study 2: Highway A20, the Netherlands}

\subsubsection{Network and Data Description}

The second case study is based on real data obtained from detectors of a stretch of the highway A20 from Rotterdam to Gouda in the Netherlands, taken from (\trbnum{Schakel2014}). The topological characteristics of this network, which incorporates a non-trivial combination of lane-drops, on-ramps, and off-ramps, the congestion pattern, and the relatively closely-spaced detectors make it a challenging test-bed for the estimation scheme proposed in the previous section.

\begin{figure}
\begin{center}
	\includegraphics[width=\textwidth]{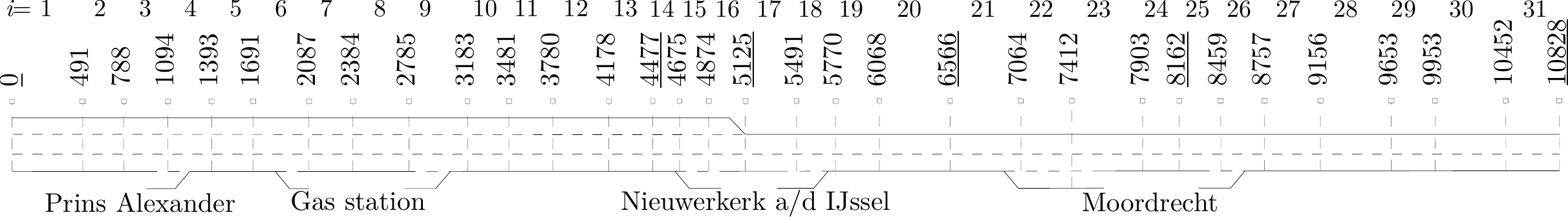}
	\caption{A graphical representation of the considered stretch of highway A20 from Rotterdam to Gouda, the Netherlands. Detector positions are indicated as the distance (in m) from the network entrance. The detectors used by the estimator for obtaining flow measurements within the case study are underlined.} 
	\label{fig:dutchStretch}
\end{center}
\end{figure}

The considered stretch is about 11\,km in length, includes 4 on-ramps and 3 off-ramps (however, the on- and off-ramp ``Gas station'' are ignored since they are characterised by extremely low flows). The stretch includes a lane-drop located at 4477\,m (see Figure \ref{fig:dutchStretch}). There are 32 detectors placed rather homogeneously at a distance of 300\,m on average, as illustrated in Figure \ref{fig:dutchStretch}. In addition, a flow detector is located at off-ramp ``Moordrecht''.

\begin{figure}
\begin{center}
	\includegraphics[width=0.9\textwidth]{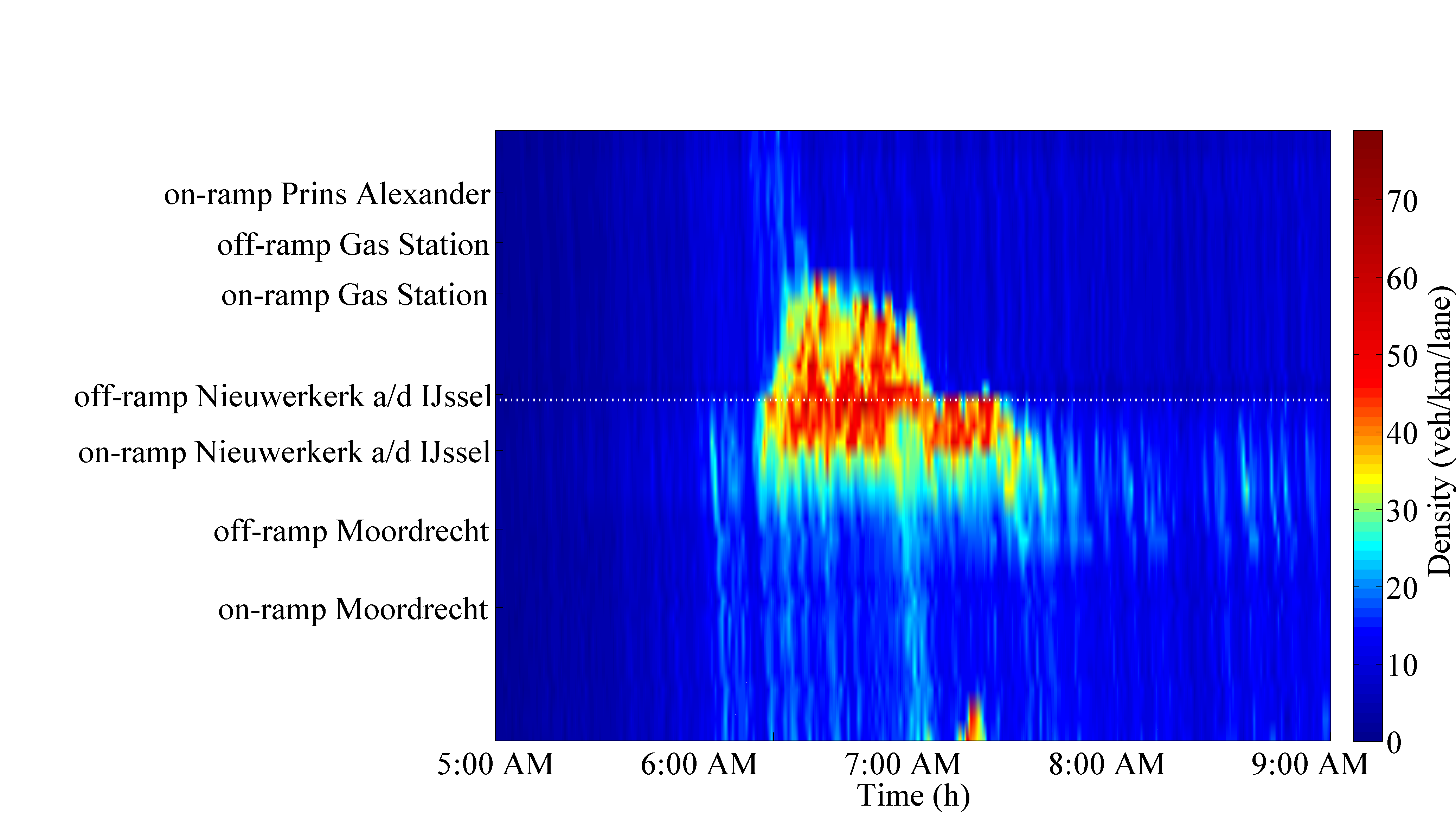}
	\caption{The space/time evolution of real densities for Case Study 2. The presence of a congestion originated at the Nieuwerkerk a/d IJssel merge area that spills-back and further deteriorate at the lane-drop, marked as a dotted line, is clearly visible.} 
	\label{fig:3d-dens_real}
\end{center}
\end{figure}

Data from the morning rush hour of Monday, June 8, 2009 are employed, where a strong congestion is created around 6:20 AM because of the increased flow entering from on-ramp ``Nieuwerkerk a/d IJssel''; consequently, the congestion spills back and worsens at the lane-drop area, reaching up to the ``Gas station'' area; then it disappears after 7:40 AM because of the reduced demand. A 3d-plot illustrating the traffic conditions from 5 AM to 9 AM is shown in Figure \ref{fig:3d-dens_real}. This congestion pattern allows to test and evaluate the proposed estimator under varying traffic conditions, which include the formation and dissipation of a stretch-internal congestion which is not visible at the stretch boundaries.

The network is space-discretised with $N=31$ segments, where each segment is delimited by a pair of detectors. According to this space-discretisaton, on-ramps are placed within segments 3, 17, and 25; whereas off-ramps are placed within segments 14 and 21 (see also Figure \ref{fig:dutchStretch}).

The proposed estimation algorithm is fed with the following information: flow measurements retrieved from a limited number of detectors according to the configuration shown in Figure \ref{fig:dutchStretch}, that is sufficient to guarantee observability as explained in the previous section (for details, see \trbcite{Bekiaris2015}); and speed measurements retrieved from all detectors, on the basis of the assumption that similar information may be obtained from connected vehicles reports. Specifically, the detector located at the highway entrance (0\,m) is used to obtain the input of the system $q_0$; all ramp flows are assumed unknown, therefore four additional detectors are utilised (one between every pair of unmeasured ramps), choosing the ones located at 4477\,m (segment $i=13$), 5125\,m (segment $i=16$), 6566\,m (segment $i=20$), and 8162\,m (segment $i=24$); and, finally, one measurement is taken at the network exit (10828\,m), employed as $q_N$ (see also Figure \ref{fig:dutchStretch}).

The ground truth is represented by the densities computed using the measurements from each detector as $\rho = q/v$, where $q$ and $v$ are the measured flow and speed respectively; and by the only ramp flow measurement available, namely the flow at off-ramp Moordrecht.

Using the same measurements both for feeding the estimator and for constructing ground truth implies that the estimator is not subject to any measurement error. However, in order to assess the performance of the estimation scheme in more realistic scenarios, two additional cases are considered. In the first, a zero-mean Gaussian white additive noise, characterised by a standard deviation of 300\,veh/h, is added to the flow measurements that are utilised by the estimator; in the second, a zero-mean Gaussian white noise, with a standard deviation of 5\,km/h, is added to the speed measurements that are employed by the filter, in addition to the flow measurement noise.

\subsubsection{Performance Evaluation}

\begin{figure}
\begin{center}
	\includegraphics[width=0.9\textwidth]{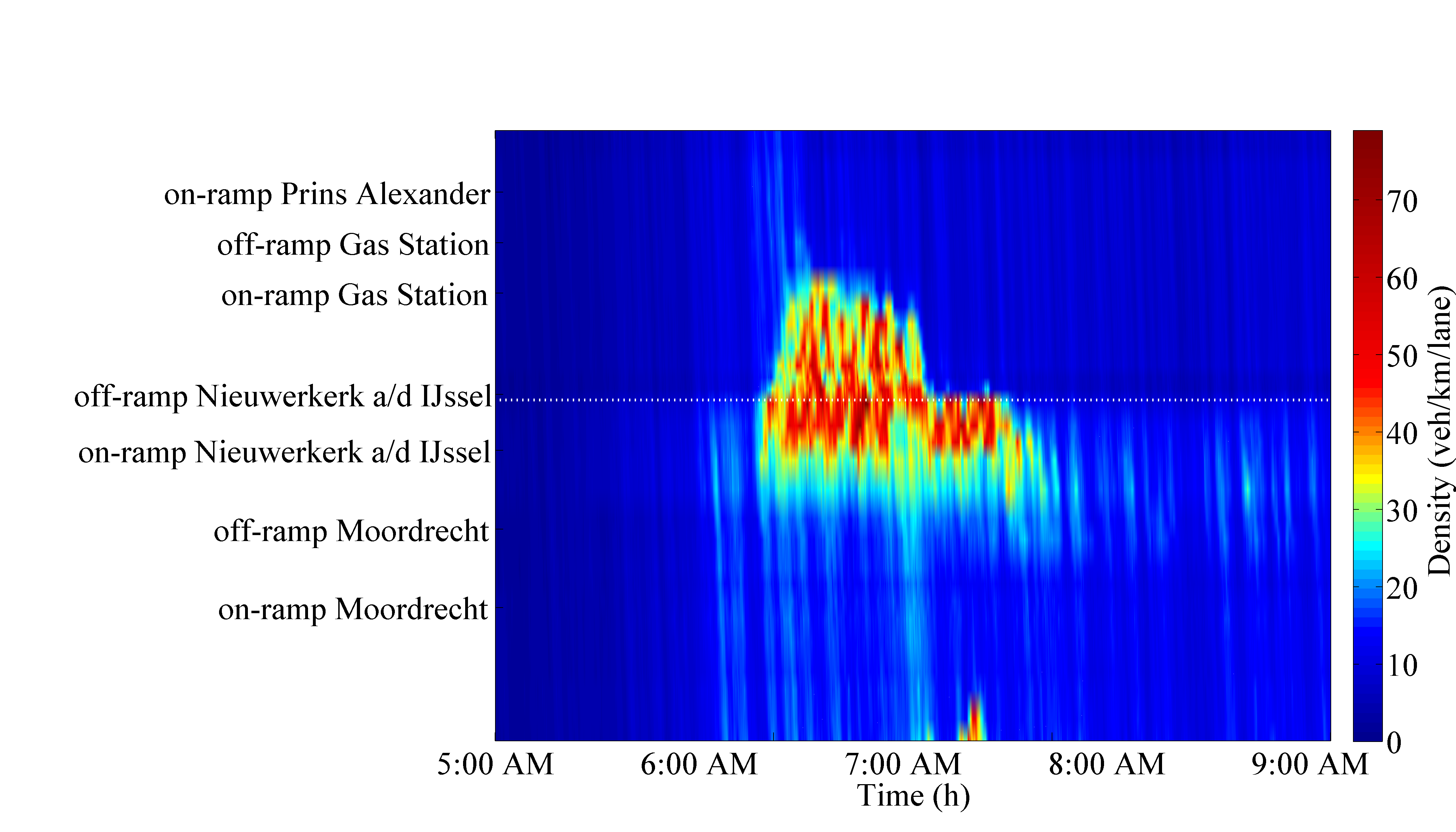}
	\caption{The space/time evolution of estimated densities for Case Study 2, when additive noise affects both flow and speed measurements. By comparison with Figure \ref{fig:3d-dens_real} it is clear that the estimator is capable of reproducing the real congestion pattern (both in time and space).}
	\label{fig:3d-dens_est}
\end{center}
\end{figure}

\begin{figure}
\begin{center}
	\includegraphics[width=\textwidth]{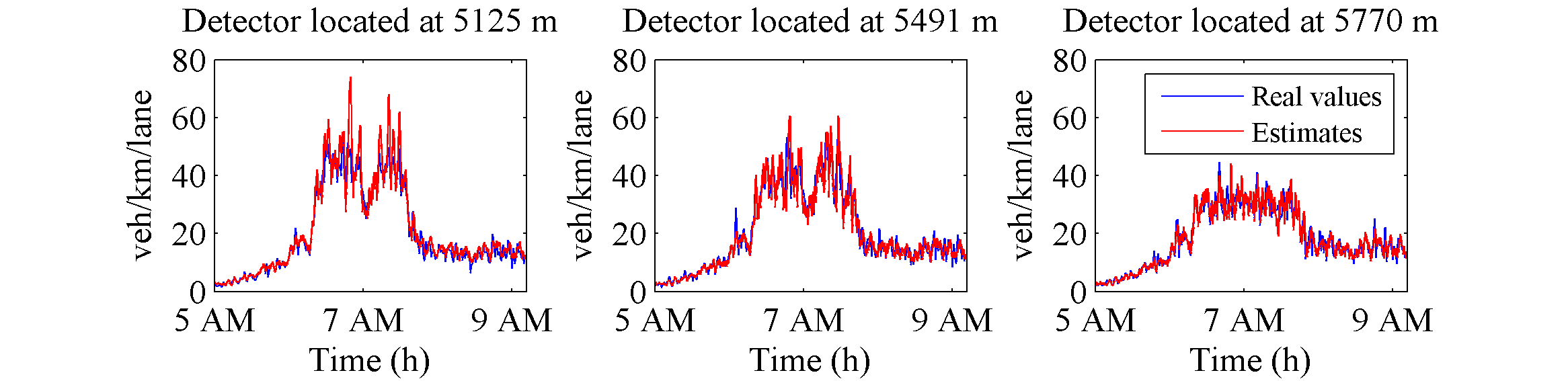}
	\caption{The real and estimated density trajectories at the location where congestion starts, for Case Study 2, when additive noise affects both flow and speed measurements. The proposed scheme is capable of estimating with high accuracy the onset of the congestion.} 
	\label{fig:trajDutch}
\end{center}
\end{figure}

\begin{figure}
\begin{center}
	\includegraphics[width=0.5\textwidth]{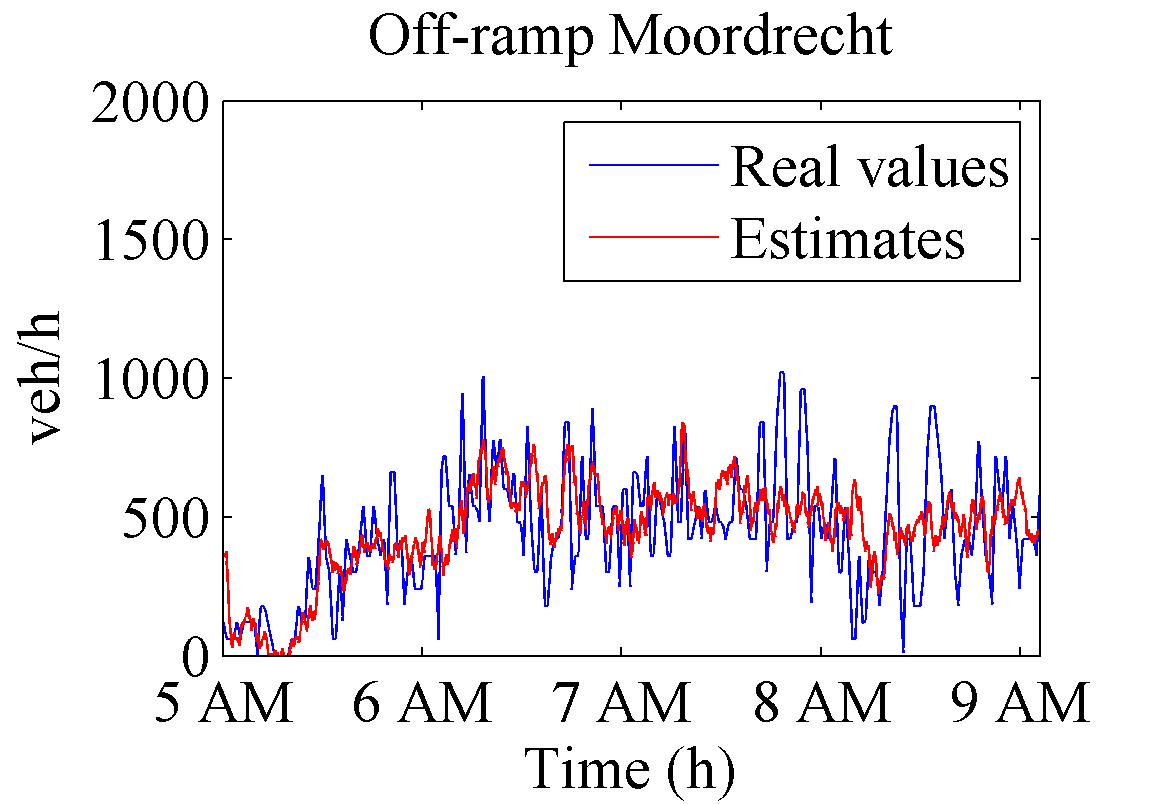}
	\caption{The measured and estimated flows at off-ramp Moordrecht, for Case Study 2,  when additive noise affects both flow and speed measurements.} 
	\label{fig:trajRampDutch}
\end{center}
\end{figure}

The parameters used for the Kalman filter are the following: $ Q_{i,i}=1, \; 0 \leq i \leq N; Q_{i,i}=0.01, \; N < i \leq N+l_r+l_s; Q_{i,j}=0, \; \forall i \neq j$;  $R=100 \; I_{ \left( l_r+l_s \right) \times \left( l_r+l_s \right) }$; $\mu = \left(4,\ldots,4\right)$; $H=I_{ \left(N+l_r+l_s\right) \times \left(N+l_r+l_s\right) }$.
The performance index in Equation \ref{eq:perfIndexDens} is used for numerical evaluation of the estimation methodology. In all the tested cases the estimates are very accurate. In the first case, where noise is not added to the measurements utilised by the estimator, a performance index value $CV_\rho=14.6\%$ is obtained. When the flow measurements used by the filter are subject to noise, we obtain $CV_\rho=15.0\%$, which is only a slight degradation with respect to the noise-free case, since the proposed estimation scheme is capable of efficiently handling the effect of additive noise. On the other hand, when noise is also added to the speed measurements, a more significant deterioration of the performance index ($CV_\rho=16.1\%$) and noisier estimates are obtained; this is due to the fact that, in this case, noise in not purely additive, but it also appears as multiplicative within the dynamic equations of the filter. By visual inspection of the 3d-plot shown in Figure \ref{fig:3d-dens_est}, which illustrates the estimated densities when noise is added to both flow and speed measurements, we can see that the estimator reliably reconstructs the congestion pattern within the network, both in time and space. Furthermore, Figure \ref{fig:trajDutch} depicts the measured and the estimated density trajectories around the location where traffic congestion starts: we can see that the estimation scheme captures the onset of congestion with accurate timing and at the correct location. Finally, Figure \ref{fig:trajRampDutch} illustrates the results of the flow estimation at off-ramp Moordrecht (the only one for which flow measurements are actually available), from which it is clear that the algorithm estimates the off-ramp flow with good accuracy.

\section{Conclusions} \label{sec:concl}

The macroscopic model-based traffic estimation scheme proposed in (\trbnum{Bekiaris2015}) has been tested on two case-studies using data stemming from different real experiments. In both cases, the results demonstrated the effectiveness of the methodology, both in qualitative and quantitative terms. The first case, based on the NGSIM microscopic data, demonstrated the capability of properly estimating the traffic state using aggregated information retrieved from moving vehicles; a set of experiments, using different penetration rates of connected vehicles, demonstrated the robustness of the proposed methodology with respect to potentially inaccurate speed measurements. The second case, where a longer highway stretch and time horizon are employed, has permitted us to verify the capability of properly estimating the traffic state in case a congestion is created within the network, making this estimation scheme particularly useful for traffic control applications. 

Ongoing work involves testing more complex scenarios, using microscopic simulations, to assess the effectiveness of the approach in case connected vehicles are characterised by a significantly different behaviour; e.g., when they are automated.

\section*{Acknowledgement}
The research leading to these results has received funding from the European Research Council under the European Union's Seventh Framework Programme (FP/2007-2013) / ERC Grant Agreement n. 321132, project TRAMAN21.

The authors would like to thank Prof. Vincenzo Punzo and Dr. Marcello Montanino from University of Naples, Italy, for their support in providing the reconstructed  NGSIM data used in the first case study; and Prof. Bart van Arem and Dr. Wouter Schakel from TU Delft, the Netherlands, for data and information related to the network used in the second case study.

\clearpage

\bibliographystyle{trb}
\bibliography{collection}

\end{document}